# Three-Dimensional Percolation Modeling of Self-Healing Composites

**Alexander Dementsov** and **Vladimir Privman**\*

Center for Advanced Materials Processing, and
Department of Physics, Clarkson University,
Potsdam, New York 13699, USA

**Abstract**

We study the self-healing process of materials with embedded "glue"-carrying cells, in the regime of the onset of the initial fatigue. Three-dimensional numerical simulations within the percolation-model approach are reported. The main numerical challenge taken up in the present work, has been to extend the calculation of the conductance to three-dimensional lattices. Our results confirm the general features of the process: The onset of the material fatigue is delayed, by developing a plateau-like time-dependence of the material quality. We demonstrate that in this low-damage regime, the changes in the conductance and thus, in similar transport/response properties of the material can be used as measures of the material quality degradation. A new feature found for three dimensions, where it is much more profound than in earlier-studied two-dimensional systems, is the competition between the healing cells. Even for low initial densities of the healing cells, they interfere with each other and reduce each other's effective healing efficiency.

___________________
\*Corresponding author

# 1. Introduction

Design of "smart materials" has been an active new research topic. Specifically, self-healing composite materials[1-5] restore their mechanical properties with time, thus reducing material fatigue caused by the formation of microcracks. Microcracks in such materials break embedded fibers/capsules[6] which contain the healing agent — a "glue" that then delays further microcrack development. In recent experiments,[1,6-11] an epoxy (polymer material) was studied, with microcapsules containing a healing agent. Cracks induce rupture of microcapsules.[1,11] Then the glue is released, permeates the crack, and a catalyst triggers re-polymerization that delays further material damage.

Nanosize defects are randomly distributed in the material: Fatigue and ultimately degradation of the material due to mechanical loads during its use, are caused by the formation of craze fibrils along which microcracks develop. Therefore, it is particularly interesting to have a self-healing process at the nanoscale: This might offer[10] a more effective prevention of growth of microcracks. It is expected[10,11] that nanoporous fibers with glue will heal smaller damage features, thus delaying the material fatigue at an earlier stage than larger capsules[1,9] which re-glue large cracks. Furthermore, on the nanoscale, the glue should be distributed/mixed with the catalyst more efficiently because transport by diffusion alone will be effective,[10] eliminating the need for external UV irradiation,[9] etc.

Presently, theoretical and numerical modeling of such a self-healing process is only in the initiation stages.[10,12-14] Theoretical works and numerical simulations[15-18] (without self-healing) have largely considered formation and propagation of large cracks which, once developed, cannot be healed by embedded nano-featured capsules. We have instead focussed[10,12] on modeling of the time dependence of a gradual formation of damage (fatigue) and its manifestation in material composition and properties, as well as its healing by nanoporous fiber rupture and release of glue.

We have formulated phenomenological rate equations[10,12] for the self-healing process. These are not reviewed here. In fact, further work is planned, with the ultimate goal of extending the rate equation approached to a full phase-field-type theory of fatigue development and self-healing, along the lines of similar approaches to crack formation.[19]

In addition to continuum modeling, numerical Monte Carlo simulations can yield useful information on the self-healing process. We reported[10,12] numerical simulations for a two-dimensional (2D) lattice model. Furthermore, calculated material composition and structure must be related to macroscopic properties that are experimentally measured. The relation between composite materials composition and properties is an important and rather broad field of research.[20]

Recently, it has been demonstrated experimentally[21] that a dilute network of carbon nanotubes, incorporated in epoxy, can provide a percolation cluster, the conductance of which reflects the degree of the fatigue of the material and also shows promise for probing the self-healing.[21] We consider conductance as a representative property. However, we point out that different transport properties can be used to probe material integrity, including thermal conductivity,[22,23] photoacoustic waves,[24,25] electrical conductivity.[21,26-29]

Transport properties can be highly nonlinear functions of the degree of damage. Specifically, the conductance can sharply drop to zero if the conducting network density drops beyond the percolation threshold. However, for probing the initial fatigue, in the regime of low levels of damage, one expects most transport properties to decrease proportionately to the damage. Our numerical 2D simulations confirmed this expectation.[12]

The purpose of the present work has been to carry out three-dimensional (3D) numerical simulations for a lattice model of (short-range[10]) self-healing of initial fatigue, along the lines of approach within the framework of a bond-percolation model, earlier developed for 2D. We point out that the main challenge was numerical. Three-dimensional simulations for percolation are numerically demanding, particularly when calculation of conductance is involved. Our simulations involved relatively large-scale simulations, as described in Section 2. In fact, to our knowledge no calculations for bond-percolation conductance for regular 3D lattices were earlier published. However, in our study we did not focus on the critical-point behavior near the percolation transition, which has been of primary interest[30-32] to the community of scientists studying percolation models per se, because we were interested in the regime of the initial damage, when the percolation network is largely intact.

Our results for the percolation model of self-healing in 3D are presented in Section 2. We find that the overall pattern of time-dependence is similar to that found earlier for 2D, for short-range self-healing. However, an interesting new finding for 3D, reported in Section 3, is that, as the initial density of the glue-carrying capsules (fibers, healing cells) is increased, they begin to interfere with each other's healing efficiency, which results in a significant suppression of the healing effect. Unlike 2D, where a similar effect is also expected, in 3D it sets in for rather low densities, well below the experimentally relevant range[1,11] of roughly 10% to 15% volume fraction of the healing cells. Section 3 addresses this property, as well as offers summarizing remarks.

## 2. Bond-Percolation Model of Self-Healing

Our Monte Carlo simulations were carried out for cubic lattices of varying sizes, $N \times N \times N$ cells, with periodic boundary conditions. One cell of a cubic lattice is shown in Figure 1. The initial fraction, $\rho$, of the lattice cells, randomly selected, were designated as glue-carrying. The kinetic rules were very similar to those used in the earlier 2D simulations[10,12] for square lattices, but with some differences specified below. Initially, at time $t = 0$, all the cell faces were intact. We then assume that as time goes by, the ongoing usage of the material causes its degradation, such that the faces of the cells break with the rate $p$.

However, if the face is that of a cell with glue in it, then the rate of its breakage is assumed larger, $P > p$. The reason for this assumption was explained in earlier works:[10,12] It represents the "cost" of embedding the healing cells, which generally should somewhat reduce the material integrity. Here we took

$$p = 0.003 \quad \text{and} \quad P = 0.008 \,. \tag{1}$$

These are convenient values representing slow rate of material degradation per single unit of time, with the dimensionless time measured as the number of Monte Carlo sweeps through the system.

Once any *two* faces (of the total 6) of a glue-carrying cell become damaged, with the second of the two "damaged" in a Monte Carlo event, then an additional process occurs. We assume that the glue in that cell immediately "leaks out," and as a result a local neighborhood of this cell is "healed" (i.e., structurally repaired). Here we used the following local-healing rule: All the 6 faces of this cell, as well as the faces of all the 26 other cells which belong to the $3 \times 3 \times 3$ cubic group of 27 cells centered at our "active" cell, are set to the "healthy"-state value. This means that irrespective of the earlier state of each of the 104 faces involved (healthy or broken) and of the state of the two respective cells that adjoin each of those faces (the cells can be regular, glue-carrying, or formerly glue-carrying but now used up), all the 104 faces are set to "healthy." This models the self-healing effect. In fact, once a face is damaged, both of its adjoining cells can become "leaky" according to this rule, because each might have had one earlier-damaged face. In this case the $3 \times 3 \times 3$ neighborhood of each is healed, which means setting all the faces in a $4 \times 3 \times 3$ box to healthy. A difference as compared to the earlier 2D simulations, was that once healed, a face of a former healing cell is later rebroken at the rate $p$, rather than $P$, unless it happen to adjoin another, not yet used, healing cell. This choice allows for a better visualization of numerical results close to $\rho = 1$, in Section 3.

The quantity measured in the simulation, after averaging over several Monte Carlo runs, was the fraction of the undamaged cells, $n(t)$, where $n(0) = 1$. This represents an "internal" measure of the material integrity as a function of time. Furthermore, as an example of a quantity measuring an "external" material property that can actually be experimentally probed, we also calculated the conductance, $G(t)$. In order to define it, let us consider a dual lattice with nodes at the centers of the original-lattice cells and with bonds, that connect these nodes, each crossing one of the original-lattice faces, as shown in Figure 1.

For our calculations, we assumed that all the bonds crossing healthy faces have the same, maximal conductance, whereas all the bonds crossing damaged faces do not conduct at all. Due to our choice of periodic boundary conditions, which reduces numerical noise, the conductance of the system was calculated between two parallel planes, each of area $N \times N$ (these were really tori, due to periodicity), at the distance $N/2$ from each other. We adapted a standard algorithm[33] to the 3D case. These two tori are connected by two equal-size system halves (we took $N$ even for simplicity), and the conductivities via these two pathways were included in the overall calculation.

Our typical results are illustrated in Figure 2, where we plot the fraction of "healthy" bonds, $n(t)$, and in Figure 3, where the conductance, $G(t)$, is shown, normalized such that initially, $G(0) = 1$. The results shown correspond to the choice of the initial healing-cell density $\rho = 1/8 = 12.5\%$. We also show the results of a numerical simulation without any healing cells. We note that is the latter case, we have

$$n(t) = e^{-pt}, \quad \text{for} \quad \rho = 0. \tag{2}$$

Since the model with $\rho = 0$ corresponds to ordinary bond percolation in 3D, the percolation transition value is known[32] from the literature: $n_{\text{percolation}}^{(\rho=0)} = 0.248\,812\,6 \pm 0.000\,000\,5$, and corresponds to time $t_{\text{percolation}}^{(\rho=0)} \approx 464$, for the value of $p$ assumed in (1).

Furthermore, since for $\rho = 0$ the bonds break without any correlation with each other, then any deviation from (2), not visible on the scale of Figure 2, as well as noise and size ($N$) dependence in the data, are only due to the statistical noise in the Monte Carlo simulation. We note that $n_{\text{percolation}}^{(\rho>0)}$ for the model with self-healing, is not known and cannot be accurately estimated from our data (because the lattice sizes are too small, see further comments below). Fortunately, interesting features of the self-healing effect occur at $n(t)$ values well over those of the percolation transition (i.e., for times much earlier than $t_{\text{percolation}}$).

With self-healing, for $\rho > 0$, there is some correlation in the bond-kinetics stochastic "history," and therefore the size dependence of $n(t)$ can be real, thought it is obviously extremely small; see Figure 2. Due to this correlation, the above value for $n_{\text{percolation}}^{(\rho=0)}$ is not valid for the self-healing case, though the deviation is expected to be small.

The data in Figures 2 and 3, and especially, Figure 4, illustrate the general features of the self-healing process. At a small initial "cost" of a faster drop in the material integrity (represented here by our model assumption that $P > p$), a plateau-like behavior is gained, at the intermediate times, resulting in an overall delay in the buildup of larger damage. For the regime of interest, of small overall degree of damage, the externally measurable material properties, exemplified here by the conductance, follow this trend and provide a reliable, approximately proportional measure of the material mechanical condition. This is illustrated in Figure 4. The correspondence breaks down once the regime of large damage, and the percolation transition, is approached.

Let us point out that the self-healing effect here is short-range and limited to the vicinity of each healing cell. Two-dimensional studies of longer-range healing yielded interesting results.[10] However, in 3D the lattice sizes amenable to simulation with present-day computer facilities, specifically if we include the calculation of the conductance, were limited to $N$ up to 12. This is further explained in the next paragraph and also commented on later. Therefore, study of longer-range self-healing in 3D was not practical. We point out that the most important difference between the present 3D study and the earlier 2D results, was the randomness in the initial placement of the healing cells, mentioned earlier. This matter will be addressed in the next section.

Our numerical simulations were parallelized on a 6-processor LINUX cluster with average CPU speeds of 2.4 GHz. The largest calculation consisted of averaging over 20 Monte Carlo runs the results for the conductance, for lattice size $N = 12$ (means a $12 \times 12 \times 12$ cubic lattice with periodic boundary conditions), and took approximately 70 CPU hours of parallel run. In fact, as mentioned in Section 1, no calculations of a 3D percolation-cluster conductance were ever published in the literature to our knowledge. The reason has been that studies of percolation are usually focused on the percolation transition point and require large lattice sizes. However, in our case the interesting self-healing effect — the delay in the sample quality degradation — occurs in the regime when the lattice is still relatively intact, away from the percolation transition. Still, it is important to check to what extent is the lattice size $12^3$ "large"? This has been the reason for us presenting the size-dependence of the results, for lattices with $N = 6, 8, 10, 12$, in Figures 2 and 3.

Due to the local nature of the self-healing rules selected, and the use of periodic boundary conditions, the size dependence of the fraction of undamaged faces (bonds, of the dual lattice) is extremely weak, and sizes $12^3$ de-facto give the infinite lattice size results, as seen in Figure 2.

Since the conductance is not a local quantity, it has a more pronounced size dependence with and without self-healing, as seen in Figure 3. As the system size grows, the conductance becomes somewhat lower. In the $N \to \infty$ limit, it approaches zero for all the time values at and larger than the percolation transition time. We did not attempt a detailed study of this critical-phenomenon type finite-size scaling dependence,[34] because, as mentioned earlier, we were interested in the behavior in the low-damage regime, well before the percolation transition times are reached. Once the material is degraded to approach the percolation-transition $n(t)$ values, it should simply be discarded in practical situations. Thus, we believe that for a relevant regime of low damage, our results up to size 12 provide a qualitatively and semi-quantitatively correct picture of the behavior of the conductance, valid in the $N \to \infty$ limit.

**3. Healing Efficiency in Three Dimensions**

The size dependence of the healthy-bond fraction was found to be negligibly small, unlike that of the conductance. In fact, if we only calculate this quantity, without the computer-resource demanding conductance, and use size $N = 10$, then we can obtain high-precision (large number of runs) evaluation of $n(t)$. We carried out such a simulation with the aim of covering the initial healing-cell fraction values up to $\rho = 100\%$. These results are shown in Figure 5.

The figure illustrates the overall pattern of behavior, with a more efficient self-healing for larger values of $\rho$. Let us define a measure of the healing efficiency by the delay-time, $\Delta t$, calculated as the displacement of the curve with $\rho > 0$ with respect to the curve with $\rho = 0$ at $n = 1/2$, along the dotted line in Figure 5. These values are plotted in Figure 6. It is obvious that the healing efficiency is not proportional to $\rho$. The cells interfere with each other and, even for rather small values of $\rho$, markedly decrease each other's healing efficiency.

The effect is quite important in 3D. To qualitatively understand this, we note that short range healing means that cells effect their "neighborhood" approximately within a distance equal to their size. Thus, in order not to interfere (literally, not to waste glue), other healing cells should not be healing this whole neighborhood, means they must be about two "shells of influence" away. The exclusion radius is thus at least three times the cell "radius," in terms of the center to center separation. It is likely even larger, depending on the specific geometry and on how strict do we want the "don't heal each other's neighborhoods" restriction to be. Let is take the less restrictive value, of three cell radii: In 2D the maximal fraction of the cells that can be placed without interference with each other is then approximately $(1/3)^2 \cong 11\%$. However, in 3D it is $(1/3)^3 \cong 3.7\%$.

The difference is that 10% (up to 15%) are typical values of the healing cell volume fractions in experiments.[1,11] Thus, in 2D the cell-interference effect is of marginal importance, and indeed earlier 2D simulations used small enough $\rho$ values to allow initial placement[10,12] of the cells sufficiently far apart (but otherwise random) to avoid all or most of the healing region overlap effects (for short-range healing).

However, in 3D the loss of healing efficiency due to overlap is much more profound. We reran the simulations that yielded the data in Figure 5, for a couple of very small values of $\rho \ll 1/27$, with initially well separated, but otherwise randomly placed, healing cells. The resulting data defined the slope of the straight line shown in Figure 6. The continuation of this straight line to larger values of $\rho$ measures the would be healing efficiency had the cells not interfered with each other. At $\rho = 1$, the loss of the efficiency is by a factor of close to 4, which is a significant effect.

Actually, our simple model of self-healing may not be fully representative of the more global effects, including the cell-cell interactions. Indeed, the self-healing effect cannot be entirely local. The glue cannot literally decompress to fill cracks in the material. In reality, it will form bridges that will relieve stress and therefore delay further growth of the advancing edges of affected microcracks. Still, our present results suggest that not just cells density but their uniform dispersion in the medium (to avoid clumping), are of importance in designing self-healing materials.

Presently, experiments with self-healing materials are very preliminary and do not yield quantitative data that could be directly modeled. Therefore, the modeling of relevance is qualitative, with the goal of guiding future experiments. For example, consideration of the offset in the formation of certain damage features, as studied in our case in terms of the delay-time, $\Delta t$, was motivated by experimental work, but for materials with embedded microcapsules[35] that did not contain "glue" (means, without the self-healing mechanism).

In summary, the present model has illustrated the overall pattern expected for the self-healing process, and also helped us uncover an interesting new, basically geometrical feature of stronger cell-cell interference in 3D as compared to 2D. However, more realistic, likely very demanding numerical model simulations are needed, as well as development of new continuum models, to further advance the theory of self-healing composites. We hope to address some of these challenges in our future work.

The authors gratefully acknowledge support of this research by the ARO under grant W911NF-05-1-0339 and by the NSF under grant DMR-0509104.

**FIGURES**

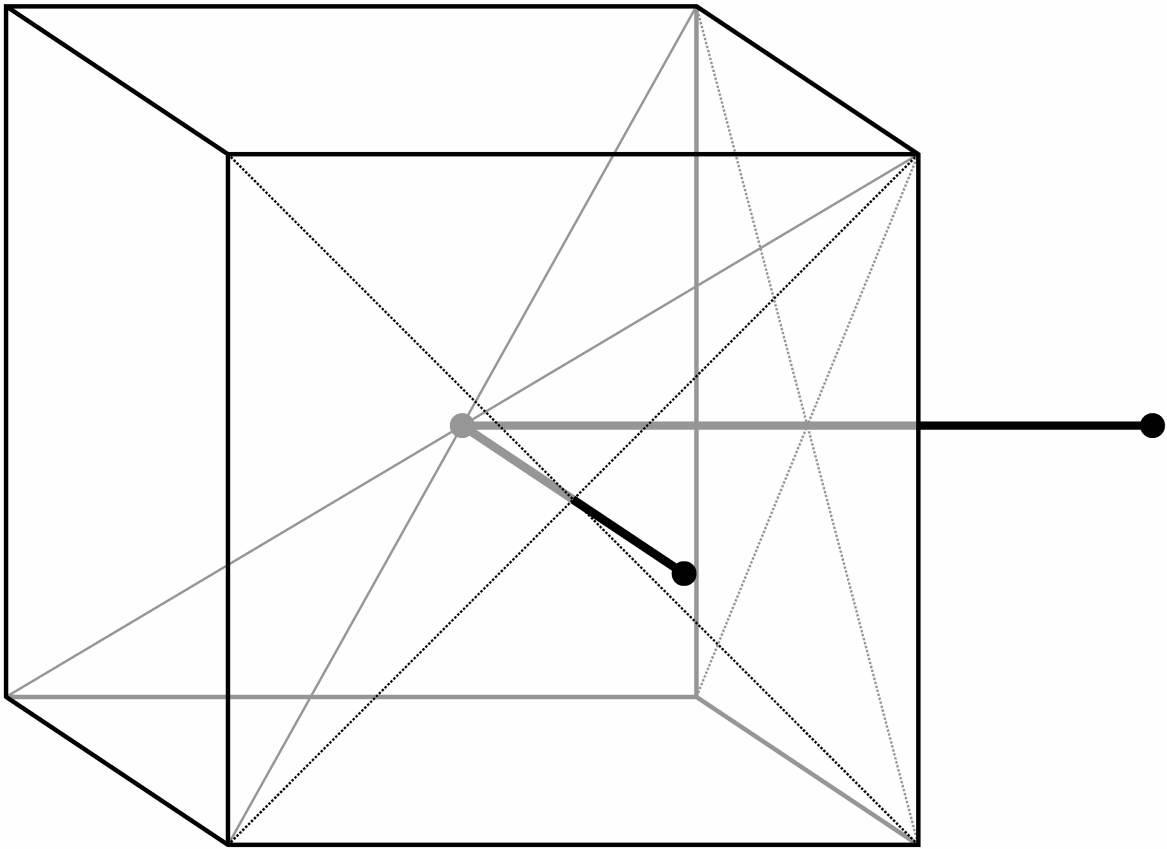

**Figure 1:** Illustration of a cell of the cubic lattice, and of two of the six bonds of the dual lattice that connect this cell's center with the centers of its nearest neighbor cells. These bonds cross the original-lattice cell faces, and we use the convention that a bond crossing an undamaged face is conducting, whereas a bond crossing a damaged face is not.

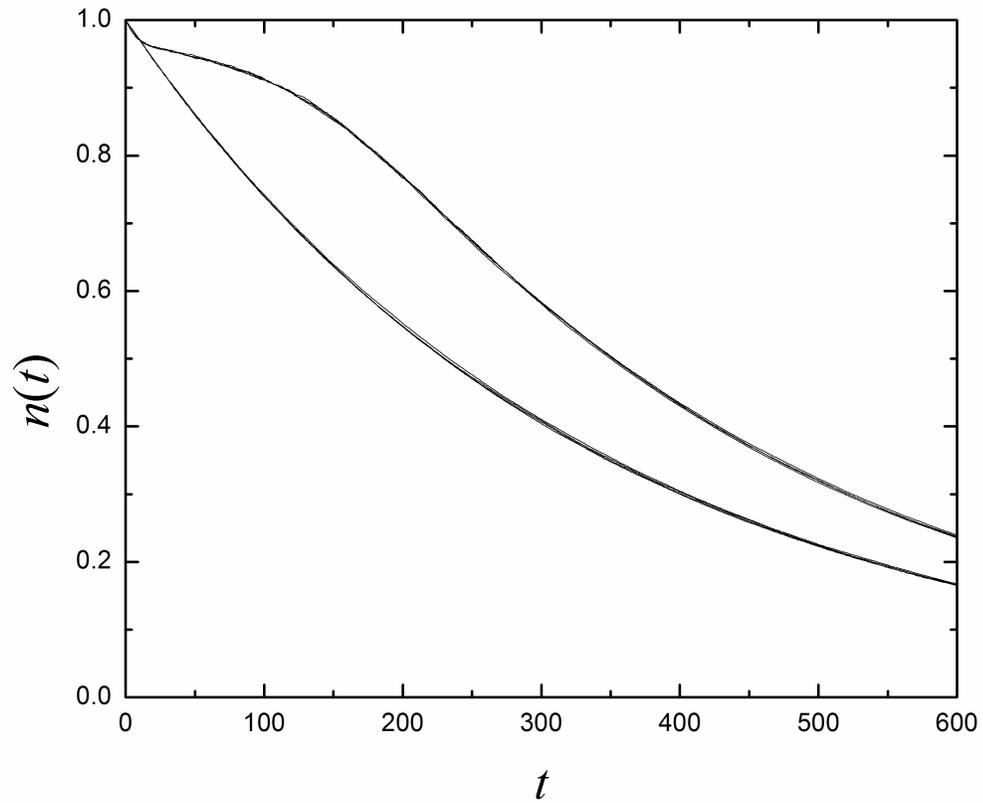

**Figure 2:** Time-dependence of the fraction of healthy bonds for the 3D lattice model. The upper (for larger times) set of data corresponds to self-healing, whereas the lower set is without self-healing. The time variable is dimensionless, measured as the number of Monte Carlo sweeps through the system. The data in each set correspond to lattice sizes $N = 6, 8, 10, 12$, but there is no measurable size dependence within the accuracy of the statistical noise. The data shown, were averaged over 240, 120, 60 and 20 Monte Carlo runs, for the four $N$ values, respectively.

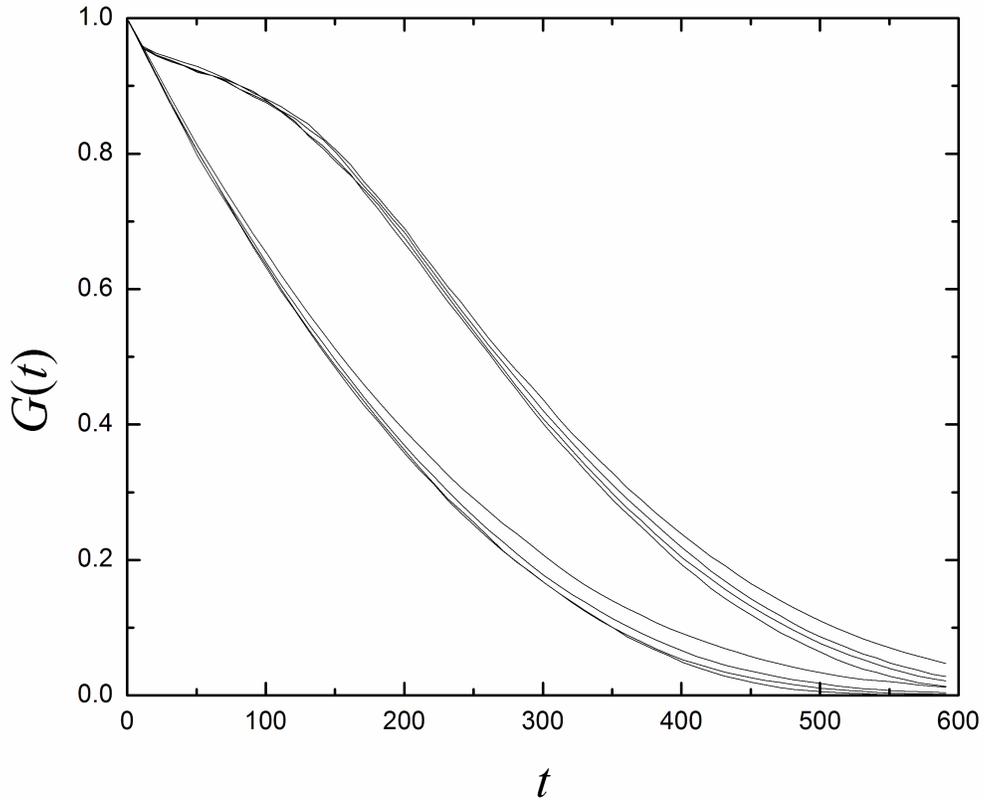

**Figure 3:** Time-dependence of the conductance for the 3D lattice model. The conductance is normalized in dimensionless units such that the value $G = 1$ corresponds to the fully connected lattice (with all the bonds conducting). The upper (for larger times) set of data corresponds to self-healing, whereas the lower set is without self-healing. The curves in each data set correspond, from top to bottom, to lattice sizes $N = 6, 8, 10, 12$. Notice the marked size dependence. The data shown, were averaged over 240, 120, 60 and 20 Monte Carlo runs, for the four $N$ values, respectively.

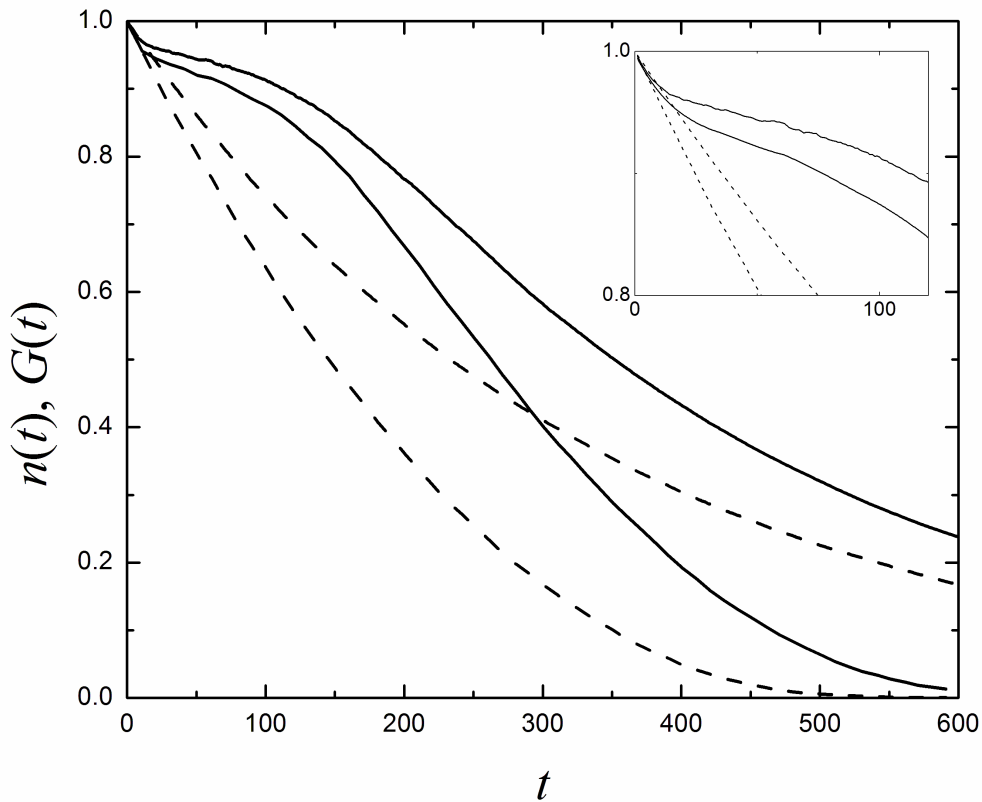

**Figure 4:** The solid curves give the fraction of the unbroken bonds, $n(t)$ — the top curve, and the normalized conductance, $G(t)$ — the bottom curve, in our 3D model with self-healing, for the initial healing-cell fraction $\rho = 12.5\%$, with the parameter values the same as in Figures 2 and 3. The data shown were obtained for the largest system size simulated, $N = 12$, averaged over 20 runs. The dashed curves illustrate similar results with all the same parameters but with no healing cells. The inset shows the data for short times, illustrating the initially slightly faster drop for the case with self-healing.

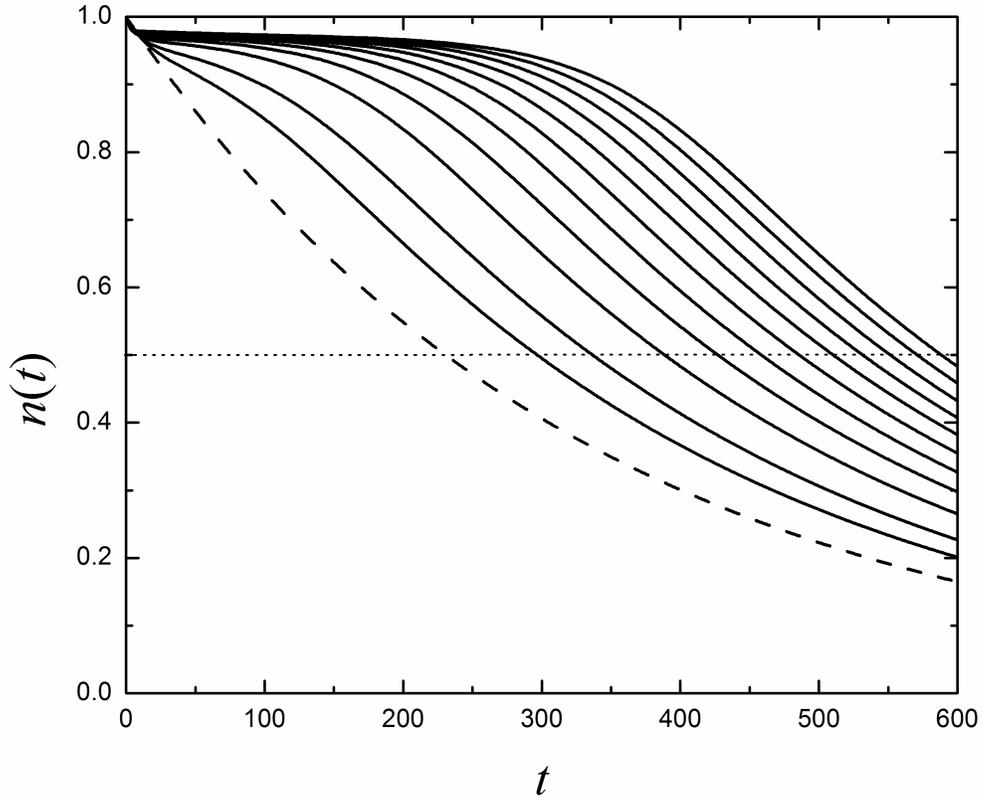

**Figure 5:** Fraction of the unbroken bonds, $n(t)$, as a function of time for the initial healing-cell densities $\rho = 5\%$, 10%, 20%, 30%, 40%, 50%, 60%, 70%, 80%, 90% and 100%. These data were obtained for $N = 10$, and each data set was averaged over 2000 runs. The dashed curve corresponds to $\rho = 0$, and increasing $\rho$ values then yielded curves with larger $n(t)$, except for the shortest times. The delay-time measure of the healing efficiency, $\Delta t$, was calculated as the displacement of the curve with $\rho > 0$ with respect to the curve with $\rho = 0$ at $n = 1/2$, i.e., along the dotted line.

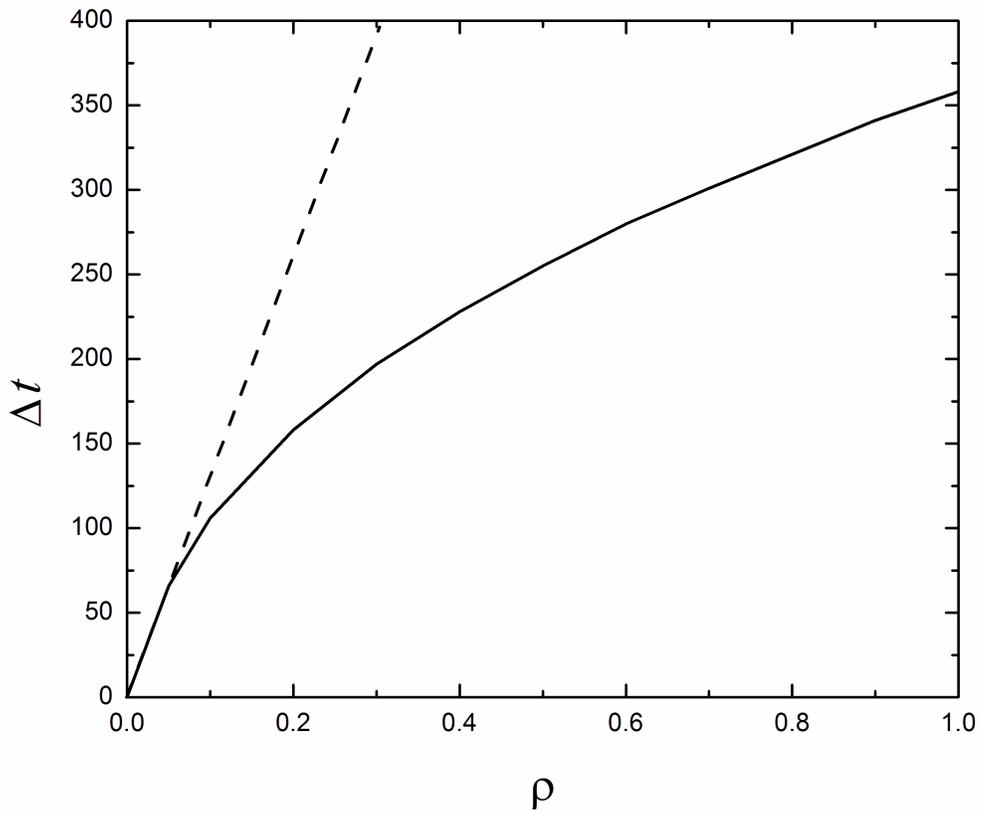

**Figure 6:** The delay-time measure of the healing efficiency, $\Delta t$, as a function of the initial healing cell density, $\rho$, extracted from the results presented in Figure 5. The solid line does not represent any data fit: it was drawn to guide the eye. The broken line represents the would be maximal healing efficiency had the cells not interfered with each other. It was obtained by a procedure described in Section 3.